\documentclass[12pt]{article}
 \usepackage{epsfig}
 \def\be{\begin{equation}}
 \def\ee{\end{equation}}
 \def\bea{\begin{eqnarray}}
 \def\eea{\end{eqnarray}}
 \usepackage{graphicx}

 \catcode`\@=11
 \def\lsim{\mathrel{\mathpalette\@versim<}}
 \def\gsim{\mathrel{\mathpalette\@versim>}}
 \def\@versim#1#2{\vcenter{\offinterlineskip
 \ialign{$\m@th#1\hfil##\hfil$\crcr#2\crcr\sim\crcr } }}
 \catcode`\@=12

 \catcode`@=12
\begin{document}
 \thispagestyle{empty}
 \begin{flushright}
 UCRHEP-T622\\
 May 2022\
 \end{flushright}
 \vspace{0.6in}
 \begin{center}
 {\LARGE \bf Nested Radiative Seesaw Masses\\ 
for Dark Matter and Neutrinos\\}
 \vspace{0.8in}
 {\bf Talal Ahmed Chowdhury\\}
 \vspace{0.1in}
{\sl Department of Physics, University of Dhaka, P.O. Box 1000, 
Dhaka, Bangladesh\\}
\vspace{0.2in}
{\bf Shaaban Khalil\\}
 \vspace{0.1in}
{\sl Center for Fundamental Physics, Zewail City of Science and Technology,\\ 
6 October City, Giza 12588, Egypt\\}
\vspace{0.2in}
{\bf Ernest Ma\\}
 \vspace{0.1in}
{\sl Department of Physics and Astronomy,\\ 
University of California, Riverside, California 92521, USA\\}
\vspace{0.8in}
\end{center}

 \begin{abstract}\
 The scotogenic model of neutrino mass is modified so that the  
 dark Majorana fermion singlet $S$ which makes the neutrino massive is itself 
 generated in one loop.  This is accomplished by having $Z_6$ lepton symmetry 
 softly broken to $Z_2$ in the scalar sector by a unique quadratic term. 
 It is shown that $S$ is a viable freeze-in dark-matter candidate through 
 Higgs decay.
 \end{abstract}

 \newpage
 \baselineskip 24pt

 \noindent \underline{\it Introduction}~:~ 
 The origin of neutrino mass~\cite{g16} may be dark matter~\cite{y17}. 
 A one-loop radiative mechanism~\cite{m06} is the possible connection, 
 known now widely as the scotogenic model.  They may also be indirectly 
 related through lepton parity~\cite{m15} or lepton number~\cite{m20}.
 There are many variants of this basic idea.  Here it is proposed that 
 the dark-matter mass itself is also radiative~\cite{mr21}.  This scenario 
 is very suitable for freeze-in light dark matter~\cite{hjmw10} through Higgs 
 decay~\cite{m19}.  It arises as the result of softly broken lepton 
 symmetry and serves as a comprehensive framework for understanding 
 neutrinos and dark matter as belonging in the same category of 
 fundamental particles.

 To implement this idea, a heavy right-handed neutrino $N$ is assumed, but it 
 is prevented from coupling directly to the left-handed neutrino $\nu$ by a 
 symmetry.  Nevertheless, both $\nu$ and $N$ couple to the dark fermion $S$. 
 The imposed symmetry is then softly broken so that $S$ gets a radiative 
 mass from $N$, and $\nu$ pairs up with $N$ in one loop through $S$. 
 This scenario is very suitable for the freeze-in mechanism 
 where the dark matter interacts very weakly and slowly builds up its relic 
 abundance, from the decay of a massive particle, in this case the Higgs boson 
 of the Standard Model (SM), before the latter itself goes out of 
 thermal equilibrium.  The direct detection of dark matter in underground 
 experiments then becomes very difficult, which is consistent 
 with the mostly null results obtained so far.

 It is well-known that baryon number $B$ and lepton number $L$ are 
 automatically conserved in the standard model (SM) of particle interactions 
 in the case of massless $\nu$.  The simplest way for it to become massive 
 is to add a singlet right-handed fermion $N_R$, then they pair up through 
 the term $\bar{N}_R (\nu_L \phi^0 - e_L \phi^+)$, where 
 $\Phi = (\phi^+,\phi^0)$ is the SM Higgs scalar doublet. This renders 
 the neutrino a Dirac mass from the vacuum expectation value 
 $\langle \phi^0 \rangle = v$, and $N_R$ is naturally assigned $L=1$.  
 On the other hand, gauge invariance also allows the $N_R N_R$ Majorana 
 mass term, hence $L$ naturally breaks to $(-1)^L$ and a seesaw mass 
 for $\nu_L$ is obtained.  In this paper, we assume $L$ to be an input 
 symmetry of the Lagrangian, so that other choices of $L$ for $N_R$ 
 are also possible~\cite{m17}. In particular, a $Z_6$ symmetry is softly 
 broken to $Z_2$.

 \noindent \underline{\it Model}~:~ 
 Each family of the SM is extended to include a right-handed fermion singlet 
 $N_R$ and a left-handed fermion singlet $S_L$.  The scalar sector consists 
 of the SM Higgs doublet $\Phi$ and a second doublet $\eta=(\eta^+,\eta^0)$ 
 together with a neutral singlet $\chi^0$.  The discrete symmetry $Z_6$ is 
 imposed on these fields as shown in Table 1, and is respected by all 
 dimension-four terms of the Lagrangian.  It is softly broken by the 
 quadratic scalar mass term $\chi^0 \chi^0$, resulting in a radiative 
 mass for $S_L$, which then induces a radiative Majorana mass for $\nu$. 
 A residual $Z_2$ discrete symmetry $D = (-1)^{L+2j}$ remains~\cite{m15}, 
 where $j$ is the intrinsic spin of the particle in question.
 \begin{table}[htb]
 \centering
 \begin{tabular}{|c|c|c|c|c|}
 \hline
 fermion/scalar & $SU(2)_L \times U(1)_Y$ & $Z_6$ & $L$ & $D$ \\ 
 \hline
 $(\nu,e)_L$ & $(2,-1/2)$ & $\omega$ & 1 & + \\ 
 $N_R$ & (1,0) & $\omega^3$ & 1 & + \\ 
 $S_L$ & (1,0) & $\omega^{-2}$ & 0 & $-$ \\ 
 \hline
 $(\phi^+,\phi^0)$ & (2,1/2) & 1 & 0 & + \\ 
 $(\eta^+,\eta^0)$ & (2,1/2) & $\omega$ & $-1$ & $-$ \\ 
 $\chi^0$ & (1,0) & $\omega$ & $-1$ & $-$ \\ 
 \hline
 \end{tabular}
 \caption{Particle content of model with $\omega^6=1$.}
 \end{table}
 
 The resulting Higgs potential is given by
 \begin{eqnarray}
 V &=& m_1^2 \Phi^\dagger \Phi + m_2^2 \eta^\dagger \eta + m_3^2 \bar{\chi}^0 
 \chi^0 + {1 \over 2} m_4^2 {\chi^0}\chi^0 + H.c. \nonumber \\ 
 &+& {1 \over 2} \lambda_1 (\Phi^\dagger \Phi)^2 + {1 \over 2} \lambda_2 
 (\eta^\dagger \eta)^2 + {1 \over 2} \lambda_3 (\bar{\chi}^0 \chi^0)^2 + 
 \lambda_{12} (\Phi^\dagger \Phi)(\eta^\dagger \eta) \nonumber \\ 
 &+& \lambda_{13}(\Phi^\dagger \Phi)(\bar{\chi}^0 \chi^0) + \lambda_{23} 
 (\eta^\dagger \eta)(\bar{\chi}^0 \chi^0) + \mu \eta^\dagger \Phi \chi^0 
 + H.c.
 \end{eqnarray}
 Let $\langle \phi^0 \rangle = v$, then the mass of the Higgs boson $H$ is 
 given by $m_H^2 = 2 \lambda_1 v^2$.  Note that the $Z_6$ symmetry is 
 respected by all dimension-four terms, but is softly broken to $Z_2$ by 
 the $m_4^2$ term.  Together with the following allowed Yukawa terms,
 \begin{equation}
 {\cal L}\supset f_\chi \bar{S}_L N_R \chi^0+f_\eta (\nu_L \eta^0 - e_L \eta^+) S_L+\mathrm{h.c},
 \label{yukawa}
 \end{equation}
 the lepton number $L$ may be assigned as shown in Table 1.  However, since the $N_R N_R$   Majorana mass term is allowed by $Z_6$, only lepton parity $(-1)^L$ is strictly 
 conserved, as is dark parity $D = (-1)^{L+2j}$~\cite{m15}.

 \noindent \underline{\it Radiative Dark Matter Mass}~:~
 The fermion singlet $N_R$ has an allowed Majorana mass $m_N$ under $Z_6$, 
 but $S_L$ is massless at tree level.  However, the breaking of $Z_6$ 
 through the soft quadratic scalar term $\chi^0 \chi^0$ allows $S_L$ 
 to acquire a radaitive Majorana mass in one loop, as shown in Fig.~1.  
 The residual symmetry of this model is then $Z_2$, which may be understood 
 as dark parity derived from lepton parity~\cite{m15}, as shown in Table 1.

 \begin{figure}[htb]
 \vspace*{-6cm}
 \hspace*{-3cm}
 \includegraphics[scale=1.0]{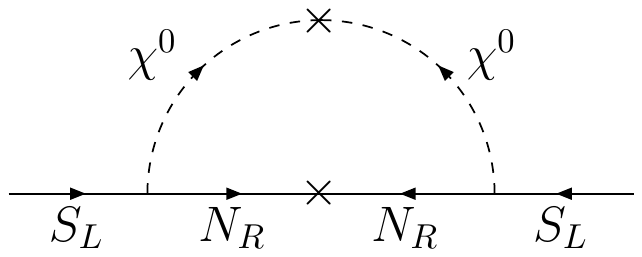}
 \vspace*{-21.5cm}
 \caption{One-loop radiative Majorana mass for the dark fermion $S$.}
 \end{figure}

 Let $\chi^0 = (\chi_R + i \chi_I)/\sqrt{2}$ and 
 $\eta^0 = (\eta_R + i \eta_I)/\sqrt{2}$, then the $2 \times 2$ 
 mass-squared matrices spanning $(\chi_R,\eta_R)$ and $(\chi_I,\eta_I)$ 
 are given by
 \begin{equation}
 {\cal M}^2_{R,I} = \pmatrix{m_3^2 + \lambda_{13} v^2 \pm m_4^2 & \mu v \cr 
 \mu v & m_2^2+\lambda_{12} v^2}.
 \end{equation}
 This means that the radiative $m_S$ comes from the difference in the 
 contributions of ${\cal M}^2_R$ and ${\cal M}^2_I$.  Let 
 $(\psi_{R1},\psi_{R2})$ be the mass eigenstates of ${\cal M}^2_R$ with 
 eigenvalues $(m^2_{R1},m^2_{R2})$:
 \begin{equation}
 \psi_{R1} = c_R \chi_R + s_R \eta_R, ~~~ \psi_{R2} = -s_R \chi_R + c_R \eta_R,
 \end{equation}
 and similarly for ${\cal M}^2_I$.  Of the 6 parameters 
 $m^2_{R1},m^2_{R2},m^2_{I1},m^2_{I2},s_R,s_I$, only 4 are independent. 
 In the limit that $m_4^2$ is very small, $s_I$ differs from $s_R$ by only 
 a small amount, say
 \begin{equation}
 s_I = s_R + \delta, ~~~ c_I = c_R - \delta (s_R/c_R),
 \end{equation}
 then
 \begin{equation}
 m_{I1}^2 = m_{R1}^2 - (\delta/s_R)(m^2_{R1}-m^2_{R2}), ~~~ 
 m^2_{I2} = m^2_{R2} - (\delta s_R/c_R^2) (m^2_{R1}-m^2_{R2}).
 \end{equation}
 Now the radiative $m_S$ mass is given by
 \begin{equation}
 m_S = {f_\chi m_N \over 32 \pi^2} [ c_R^2 F(m^2_{R1},m_N^2) - c_I^2 
 F(m^2_{I1},m_N^2) + s_R^2 F(m_{R2}^2,m_N^2) - s_I^2 F(m^2_{I2},m_N^2)]f^{T}_{\chi},
 \label{mSeq}
 \end{equation}
 where $F(a,b) = a \ln(a/b)/(a-b)$ and $f_\chi$ is the $\bar{S}_L N_R \chi^0$ 
 coupling.

 \noindent \underline{\it Radiative Neutrino Mass}~:~
 Since $S_L$ gets a radiative mass, $\nu_L$ is now connected to $N_R$ as 
 shown in Fig.~2.  Call this Dirac mass $m_D$, then the neutrino gets 
 the usual seesaw Majorana mass $m_D^2/m_N$.

 \begin{figure}[htb]
 \vspace*{-5cm}
 \hspace*{-3cm}
 \includegraphics[scale=1.0]{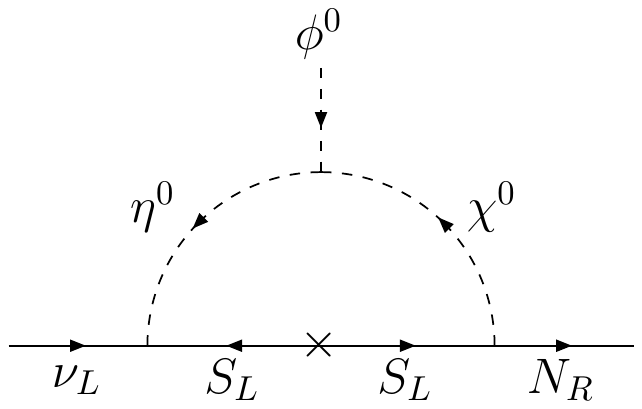}
 \vspace*{-21.5cm}
 \caption{One-loop radiative Dirac mass linking $\nu_L$ to $N_R$.}
 \end{figure}

 Since $m_S$ itself is suppressed by $m_N^{-1}$ from Fig.~1, $m_\nu$ gets 
 suppressed by $m_N^{-3}$ in this case.  On the other hand, there is 
 a diagram for Majorana $m_\nu$ directly as shown in Fig.~3, which is 
 suppressed by only $m_N^{-1}$.

 \begin{figure}[htb]
 \vspace*{-5cm}
 \hspace*{-3cm}
 \includegraphics[scale=1.0]{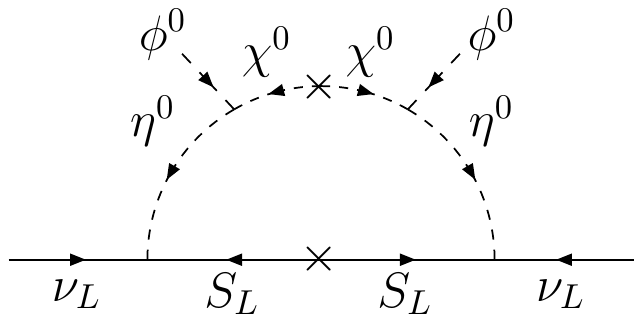}
 \vspace*{-21.5cm}
 \caption{Scotogenic Majorana mass for $\nu$.}
 \end{figure}

 The radiative neutrino mass $m_\nu$ is then generated from $m_S$ in exact 
 analogy to $m_S$ from $m_N$, i.e.
 \begin{eqnarray}
 m_\nu &=& {f_\eta m_S \over 32 \pi^2} [ s_R^2 F(m^2_{R1},m_S^2) - s_I^2 
 F(m^2_{I1},m_S^2) + c_R^2 F(m_{R2}^2,m_S^2) - c_I^2 F(m^2_{I2},m_S^2)]f^{T}_{\eta},\nonumber\\
 &=& f_{\eta}.\Lambda.f^{T}_{\eta}
 \label{mnueq}
\end{eqnarray}
where $f_\eta$ is the $\nu_L S_L \eta^0$ coupling and $\Lambda$ is the diagonal matrix containing the loop functions.  Just as $m_S$ is a 
function of $f_\chi$, $s_L$, $\delta$, $m_{R1}$, $m_{R2}$, and $m_N$, 
$m_\nu$ is a function of $f_\eta$, $s_L$, $\delta$, $m_{R1}$, $m_{R2}$, 
and $m_S$.  We explore below the possible parameter space for the dark 
matter mass $m_S$ and the neutrino mass $m_\nu$.

\noindent \underline{\it The Viable Parameter Space}~:~ 
For a general complex Yukawa $f_{\chi}$, the mass eigenstates for the $S$ fermions with corresponding mass eigenvalues can be determined from Eq.~\ref{mSeq} by following the diagonalization procedure of a complex symmetric matrix. For simplicity, we consider the Yukawa matrix to be $f_{\chi}=\mathrm{diag}(1,1,1)$, and the lightest fermion $S_{1}$ is considered as the DM candidate. Besides, the masses of $S_{1,2,3}$ depend on the RH neutrino masses $m_{N_{1,2,3}}$ and the scalar masses $m_{R_{i}}$, $m_{I_{i}}$, $s_{R}$ and $\delta$.
\begin{figure}[htb]
\centerline{\includegraphics[width=10cm]{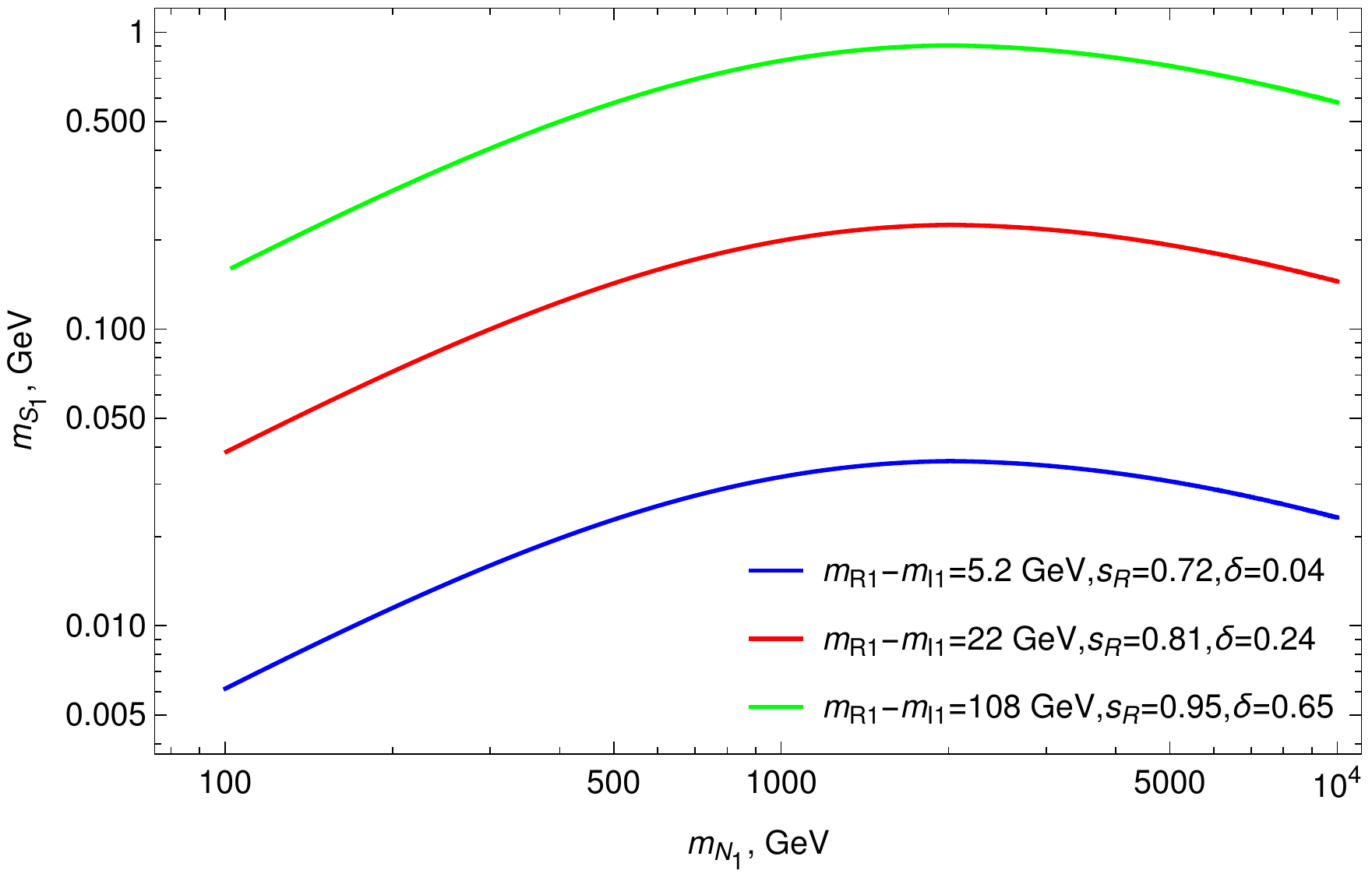}}
\caption{Correlation between the lightest RH neutrino mass, $m_{N_1}$ and the DM mass, $m_{S_{1}}$ for three fixed mass separations, $m_{R1}-m_{I1}$, $s_{R}$ and $\delta$. Here, $m_{N_{2}}=m_{N_{1}}+100$ GeV and $m_{N_{3}}=m_{N_{2}}+200$ GeV. Moreover, the mass parameters $m_{2}$ and $m_{3}$ are fixed at $m_{2}=m_{3}=1000$ GeV, and the cubic term, $\mu=500$ GeV.}
\label{mN1vsmS1fig}
\end{figure}
As shown in Fig.~\ref{mN1vsmS1fig}, the mass of the DM candidate $S_{1}$ increases if the mass splitting between $\psi_{R}$ and $\psi_{I}$ increases, which in turn depends on the larger value of the mass parameter $m_{4}$, associated with the quadratic scalar mass term that softly breaks $Z_{6}$ to $Z_{2}$, resulting in the radiative masses for $S_{i}$.

Besides, the Yukawa matrix $f_{\eta}$ can be written using the Casas-Ibarra parametrization \cite{Casas:2001sr} in the following way,
\begin{equation}
f_\eta=U_{\mathrm{PMNS}}\,\sqrt{\hat{m}}\,R\,\sqrt{\Lambda^{-1}}
\end{equation}
where $U_{\mathrm{PMNS}}$ is the PMNS matrix of neutrino mixing, $\hat{m}=\mathrm{diag}(m_{1},m_{2},m_{3})$ shows the neutrino masses, and $R$ is a complex orthogonal matrix whose angles are taken to be real in our case for simplicity. In addition, the Yukawa term involving $f_{\eta}$ coupling leads to the charged lepton flavor violation (LFV) for our model. We consider the constraints on the following charged LFV processes: $\mathrm{Br}(\mu\rightarrow e\gamma)<4.2\times 10^{-13}$ \cite{MEG:2016leq}, $\mathrm{Br}(\tau\rightarrow e\gamma)<5.6\times 10^{-8}$ and $\mathrm{Br}(\tau\rightarrow \mu\gamma)<4.2\times 10^{-8}$ \cite{Belle:2021ysv}, $\mathrm{Br}(\mu\rightarrow 3e)< 10^{-12}$ \cite{SINDRUM:1987nra}, $\mathrm{Br}(\tau\rightarrow 3 e)<2.7\times 10^{-8}$ and $\mathrm{Br}(\tau\rightarrow 3 \mu)< 2.1\times 10^{-8}$ \cite{Hayasaka:2010np}, $\mu-e$ conversion in Ti $<1.7\times 10^{-12}$ \cite{SINDRUMII:1998mwd} and $\mu-e$ conversion in Au $<7\times 10^{-13}$ \cite{SINDRUMII:2006dvw} to determine the viable region of the parameter space with our numerical analysis.

\begin{figure}[htb]
\vspace*{-5cm}
\hspace*{-3cm}
\includegraphics[scale=1.0]{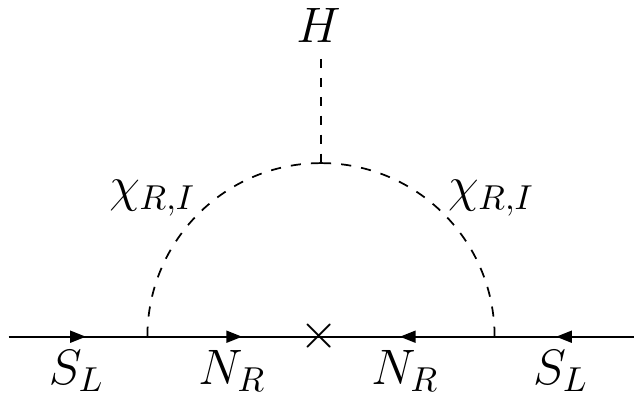}
\vspace*{-21.5cm}
\caption{One-loop diagram of the interaction among the Higgs boson and the dark fermions, $S$.}
\label{higgsSfig}
\end{figure}

\noindent \underline{\it Production of the Dark Fermion $S$}~:~
The dark fermions $S$ can have the effective interaction $f_{h}\overline{S}Sh$ with the Higgs boson generated at one loop shown in Fig.~\ref{higgsSfig},
where the relevant Yukawa coupling $f_{h}$ is given by
\begin{equation}
f_{h}=\frac{\lambda_{13} v}{32\pi^2}f_{\chi}m_{N}\left[c^{2}_{R}G(m_{R1}^{2},m_{N}^2)-c^{2}_{I}G(m_{I1}^{2},m_{N}^2)+s^{2}_{R}G(m_{R2}^{2},m_{N}^2)-s^{2}_{I}G(m_{R1}^{2},m_{N}^2)\right]f^{T}_{\chi},\label{fhcoup}
\end{equation}
with $G(a,b)=1/(a-b)-b\,\mathrm{ln}(a/b)/(a-b)^{2}$.

Based on this effective interaction between the Higgs and the dark matter candidate $S_1$, one can consider the freeze-in mechanism \cite{hjmw10,McDonald:2001vt}  to achieve the correct DM relic abundance with the following considerations.
\begin{itemize}
\item The reheating tempeature has been set as $T_{R}\ll m_{N_{1,2,3}},\,m_{R1,R2},\,m_{I1,I2},\,m_{\eta^{+}}$ so that during the gradual increase of the abundance of FIMP, $S_{1}$ from an initially negligible  value at the early universe (i.e. at $T_R$) through the decay and scattering from the thermal bath particles, the abundances of $N_{i}$, $\psi_{R}$ and $\psi_{I}$ are already Boltzmann suppressed as their presence in the thermal bath would led to excessive relic abundance of $S_{1}$ via the processes, controlled by the Yukawa couplings, $f_{\eta}$ and $f_{\chi}$, noted below, 
\begin{equation}
\overline{S}_{1}S_{1}\leftrightarrow\overline{N_{i}}N_{j},\, \psi_{R_{i}}\psi_{R_{j}},\,\psi_{I_{i}}\psi_{I_{j}},\,\psi_{R_{i}}\psi_{I_{j}},\,\eta^{+}\eta^{-},\,\overline{\nu}\nu,\,l^{+}l^{-}\nonumber
\end{equation}
Therefore, $T_{R}$ is set at the $T_{R}\sim T_{c}=159.5$ GeV \cite{DOnofrio:2015gop}, which is the Standard Model cross-over temperature.

\item Now the relevant processes which contribute to the freeze-in of the DM are the decay of the Higgs into dark fermions, $S_{1}$, and the scattering of the SM fermions $f$ and gauge bosons $V$ via Higgs boson at $s$-channel as follows,
\begin{equation}
h\rightarrow \overline{S_{1}}S_{1},\,\,\,\overline{f}\,f\rightarrow \overline{S_{1}}S_{1},\,\,\,V\,V\rightarrow \overline{S_{1}}S_{1}
\nonumber
\end{equation}
\end{itemize}

\begin{figure}
\centerline{\includegraphics[width=9cm]{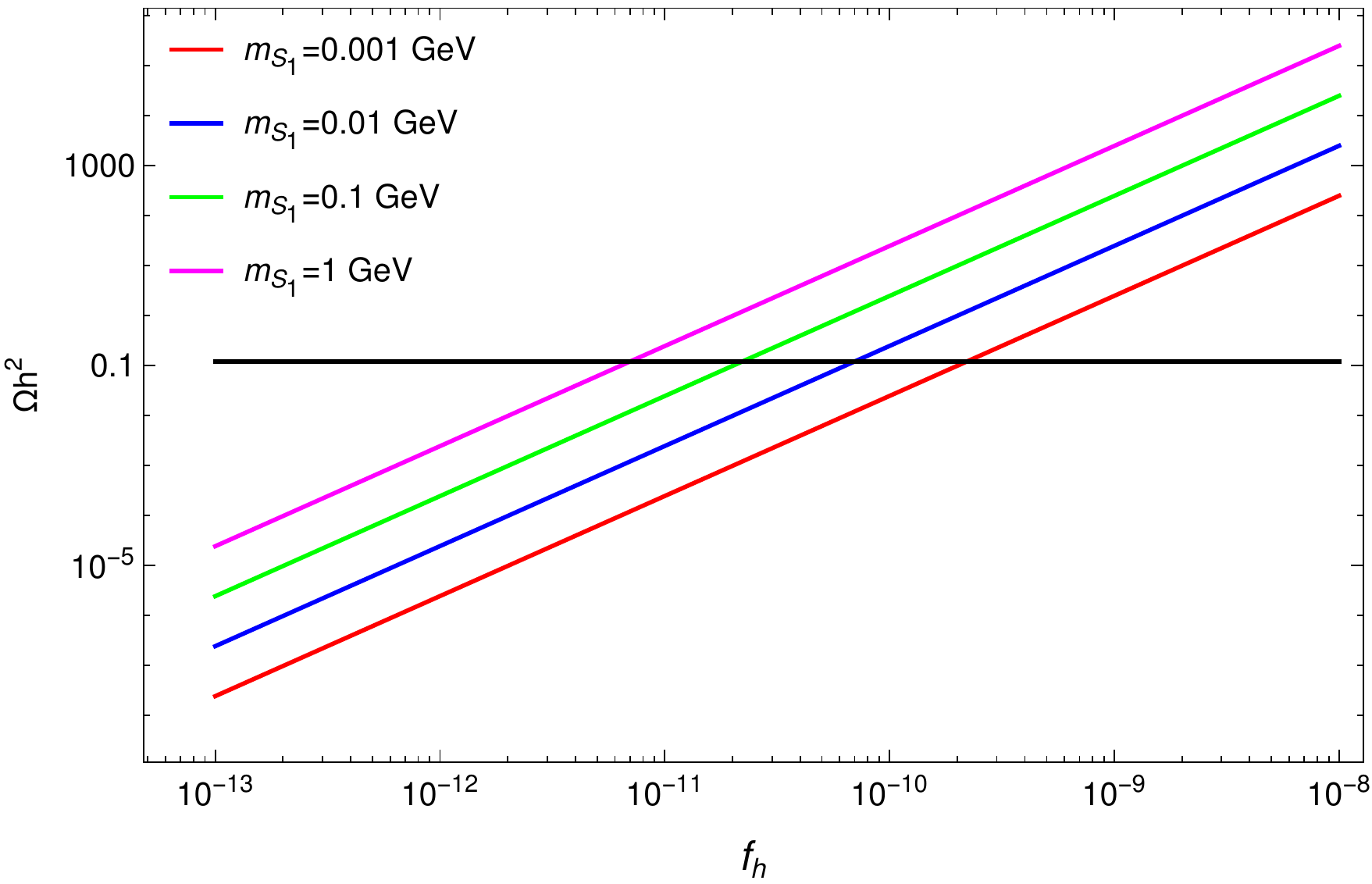}\hspace{1mm}\includegraphics[width=9cm]{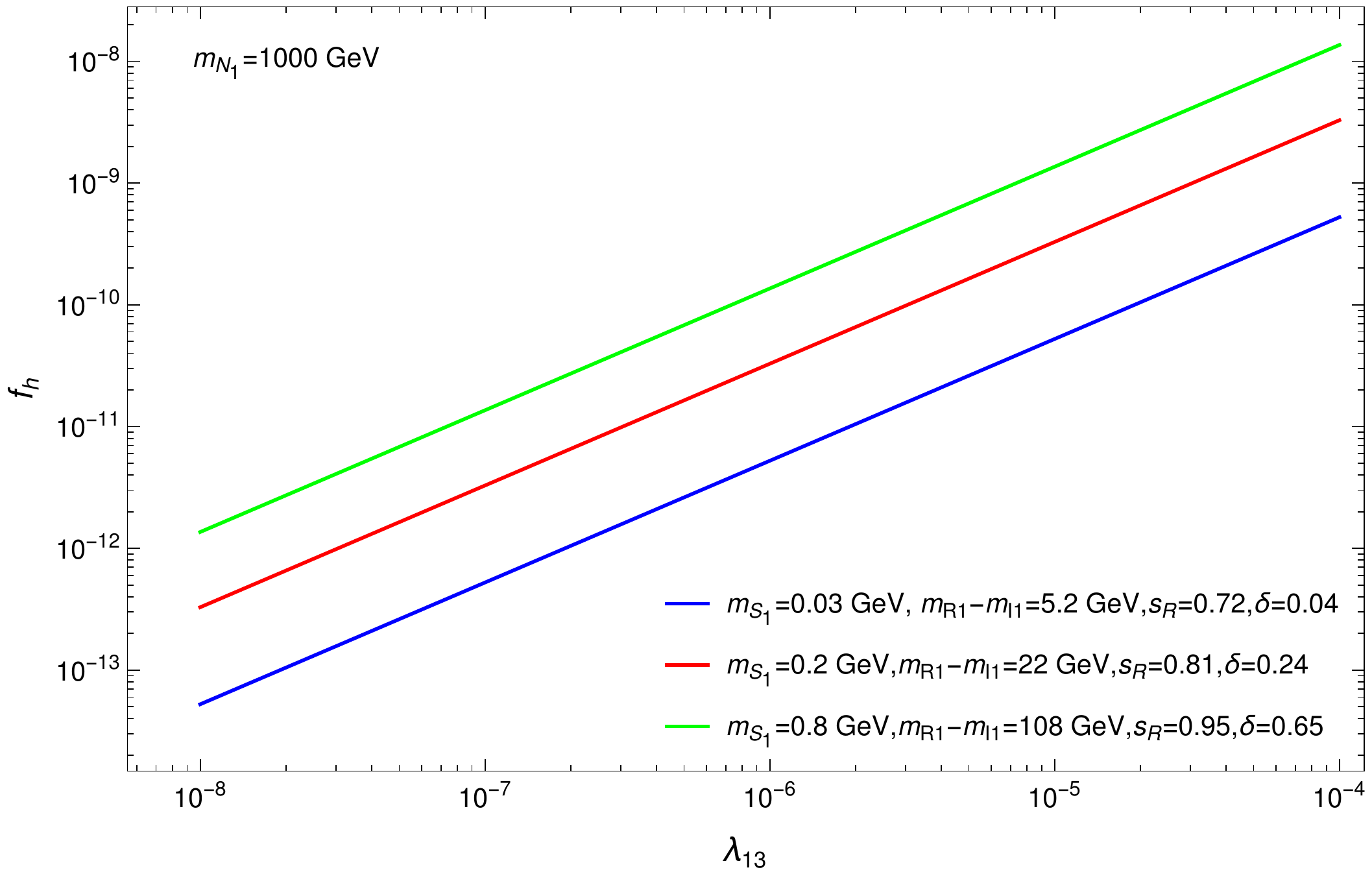}}
\caption{(left) DM relic abundance with respect to the effective Higgs-DM coupling $f_{h}$ for different DM masses via the freeze-in mechanism. The horizontal black line represent the observed DM relic abundance, $\Omega h^{2}=0.12\pm 0.001$ ($68\%$ C.L) \cite{Planck:2018vyg}. (Right) Correlation between the couplings $\lambda_{13}$ and $f_{h}$ for fixed DM mass, $m_{N_{1}}$, $m_{R1}-m_{I1}$, $s_{R}$ and $\delta$.}
\label{freezeinfig}
\end{figure}
We calculate the relic abundance following the formalism presented in \cite{Belanger:2018ccd}. We find out that for the DM mass varying from $0.001$ GeV to $1$ GeV, the required value of the $f_{h}$ coupling is $2\times 10^{-10}$ to $8\times 10^{-12}$, respectively, as seen from Fig.~\ref{freezeinfig} (left). Besides, we can see from Fig.~\ref{freezeinfig} (right) that for a fixed value of the DM mass or in other words for fixed values of $m_{N_{1}}$, $m_{R1}-m_{I1}$, $s_{R}$ and $\delta$, one can adjust the value of the scalar coupling $\lambda_{13}$ to achieve the $f_{h}$ value for the correct relic abundance.

\noindent \underline{\it Conclusion}~:~ In this work, we address a modification of the scotogenic model of neutrino mass where the mass of the fermionic dark matter itself is radiatively generated in one loop.  We determine the viable parameter space which satisfies the current limits on the charged lepton flavor violating processes.  Within that parameter space, we then calculate the DM relic abundance in the freeze-in mechanism through the decay of the Higgs boson and the $2\rightarrow 2$ scatterings of SM fermions and gauge bosons into DM pairs. We find that for the light DM in our model with mass ranging from $m_{S_{1}}=0.001$ GeV to $m_{S_{1}}=1$ GeV, the required effective coupling $f_h$ between the Higgs and the DM generated at one-loop has to be from $f_{h}=2\times 10^{-10}$ to $f_{h}=8\times 10^{-12}$ respectively, to obtain the correct relic abundance. Increasing the value of the DM mass thus requires smaller value of $f_{h}$. Furthermore, we observe that such loop-suppressed small values of the effective coupling $f_{h}$ can be achieved for the DM mass range, $m_{S_{1}}=0.001-1$ GeV by adjusting the scalar quartic coupling $\lambda_{13}$, which turns out be relatively larger than the $f_{h}$ and within the range $O(10^{-8}-10^{-6})$.

\noindent \underline{\it Acknowledgement}~:~
The work of E.M. was supported in part by the U.~S.~Department of Energy Grant 
No. DE-SC0008541. Also, S.K. acknowledges support from Science, Technology $\&$ Innovation Funding Authority (STDF) Egypt, under grant number 37272.

\bibliographystyle{unsrt}

\begin{thebibliography}{99}
\bibitem{g16} See for example A. de Gouvea, Ann. Rev. Nucl. Part. Sci. 
{\bf 66}, 197 (2016).
\bibitem{y17} See for example B.-L. Young, Front. Phys. {\bf 12}, 121201, 
121202 (2017).
\bibitem{m06} E. Ma, Phys. Rev. {\bf D73}, 077301 (2006).
\bibitem{m15} E. Ma, Phys. Rev. Lett. {\bf 115}, 011801 (2015).
\bibitem{m20} E. Ma, Phys. Lett. {\bf B809}, 135736 (2020).
\bibitem{mr21} E. Ma and V. De Romeri, Phys. Rev. {\bf D104}, 055004 (2021).
\bibitem{hjmw10} L. J. Hall, K. Jedamzik, J. March-Russell, and S. M. West, 
JHEP {\bf 1003}, 080 (2010).
\bibitem{m19} E. Ma, LHEP {\bf 2} (1) 103 (2019) [arXiv:1810.06506 [hep-ph]].
\bibitem{m17} E. Ma, Mod. Phys. Lett. {\bf A32}, 173007 (2017).

\bibitem{Casas:2001sr}
J.~A.~Casas and A.~Ibarra,
Nucl. Phys. B \textbf{618} (2001), 171-204
\bibitem{MEG:2016leq}
A.~M.~Baldini \textit{et al.} [MEG],
Eur. Phys. J. C \textbf{76} (2016) no.8, 434

\bibitem{Belle:2021ysv}
A.~Abdesselam \textit{et al.} [Belle],
JHEP \textbf{10} (2021), 19

\bibitem{SINDRUM:1987nra}
U.~Bellgardt \textit{et al.} [SINDRUM],
Nucl. Phys. B \textbf{299} (1988), 1-6

\bibitem{Hayasaka:2010np}
K.~Hayasaka, K.~Inami, Y.~Miyazaki, K.~Arinstein, V.~Aulchenko, T.~Aushev, A.~M.~Bakich, A.~Bay, K.~Belous and V.~Bhardwaj, \textit{et al.}
Phys. Lett. B \textbf{687} (2010), 139-143

\bibitem{SINDRUMII:1998mwd}
J.~Kaulard \textit{et al.} [SINDRUM II],
Phys. Lett. B \textbf{422} (1998), 334-338

\bibitem{SINDRUMII:2006dvw}
W.~H.~Bertl \textit{et al.} [SINDRUM II],
Eur. Phys. J. C \textbf{47} (2006), 337-346

\bibitem{McDonald:2001vt}
J.~McDonald,
Phys. Rev. Lett. \textbf{88} (2002), 091304

\bibitem{DOnofrio:2015gop}
M.~D'Onofrio and K.~Rummukainen,
Phys. Rev. D \textbf{93} (2016) no.2, 025003

\bibitem{Belanger:2018ccd}
G.~B\'elanger, F.~Boudjema, A.~Goudelis, A.~Pukhov and B.~Zaldivar,
Comput. Phys. Commun. \textbf{231} (2018), 173-186

\bibitem{Planck:2018vyg}
N.~Aghanim \textit{et al.} [Planck],
Astron. Astrophys. \textbf{641} (2020), A6
[erratum: Astron. Astrophys. \textbf{652} (2021), C4]


\end{thebibliography}

\end{document}